\documentclass[twocolumn,prb]{revtex4}
\usepackage{graphicx}
\usepackage{dcolumn}
\usepackage{bm}
\usepackage{amsmath}

\begin{document}

\preprint{}
\title{ Microscopic theory of flexo Dzyaloshinskii-Moriya-type interaction }
\author{Takehito Yokoyama}
\affiliation{Department of Physics, Institute of Science Tokyo, Tokyo 152-8551,
Japan 
}
\date{\today}

\begin{abstract}
We study interaction between two magnetic impurities mediated by itinerant electrons on the surface of curved magnets based on perturbation theory.
We show that Dzyaloshinskii-Moriya type interaction can arise from inhomogeneous spin texture by bending, without any spin-orbit coupling. 
Analytical expressions of the Dzyaloshinskii-Moriya type interaction are obtained. We demonstrate this effect in a one-dimensional ring model.

\end{abstract}

\maketitle


\section{Introduction}

The flexoelectric effect--namely, the linear coupling between electric polarization and strain gradients--constitutes a universal electromechanical response permitted by symmetry in all insulating solids, irrespective of crystallographic point group. \cite{Zubko2013,Yudin2013,Wang2019} In contrast to piezoelectricity, which requires broken inversion symmetry, flexoelectricity arises from the nonuniformity of mechanical deformation and is therefore present even in centrosymmetric materials. This fundamental distinction renders flexoelectricity particularly relevant at reduced length scales, where strain gradients can become extremely large and can be deliberately engineered. As a result, flexoelectricity has been extensively explored in thin films, nanostructures, and membranes, including two-dimensional crystalline and biological systems, in which curvature and bending naturally generate strain gradients.\cite{Ahmadpoor2015}
Recent studies have revealed diverse manifestations of flexoelectricity in low-dimensional and geometrically constrained systems. For example, flexoelectric coupling to antiferrodistortive modes can destabilize homogeneous phases and induce spatially modulated structures in multiferroics.\cite{Eliseev2013}
Giant flexoelectric responses driven by interfacial strain relaxation have been observed in oxide epitaxial thin films,\cite{LeeNoh2012} and flexoelectric effects have also been reported in centrosymmetric semiconductors.\cite{Wang2020}
Moreover, the emergence and transformation of polar skyrmion lattices mediated by flexoelectricity have been discussed.\cite{Ren2024}
The capability to engineer large strain gradients in nanostructures has further expanded the design space for flexo-functional materials.\cite{Tang2017,XiaGuo2019}
From a theoretical perspective, the flexoelectric effect is described by a fourth-rank tensor that couples electric polarization to strain gradients and has been systematically investigated within the framework of equilibrium thermodynamics.\cite{FuCross2007}

Microscopically, the flexoelectric tensor can be decomposed into ionic and electronic contributions. The ionic contribution arises from relative displacements of atomic sublattices under nonuniform strain fields, whereas the electronic contribution originates from strain-gradient-induced redistribution of the electronic charge density.
The relationship between flexoelectricity and bulk electric multipole moments has been elucidated.\cite{Resta2010}
Furthermore, density functional theory has been formulated to enable first-principles calculations of flexoelectric tensors.\cite{HongVanderbilt2011,Stengel2016}

In parallel, flexoelectricity has emerged as a key ingredient in a wide range of strain-gradient-driven phenomena beyond polarization, including phonon-related effects,\cite{Morozovska2015,Morozovska2025} metallic polarity,\cite{ZabaloStengel2021,YurkovYudin2021,Peng2024} and flexophotovoltaic responses.\cite{Yang2018,Guo2020,Jiang2021,Wang2024}

The concept of flexoelectricity has been generalized far beyond polarization, encompassing analogous couplings between strain gradients and other order parameters.\cite{Bukharaev2018,Du2023} In magnetic systems, strain-gradient-induced modulation of exchange interactions, magnetic anisotropy, or Dzyaloshinskii-Moriya interactions gives rise to flexomagnetism and spin flexoelectricity.\cite{Tang2025} It has been theoretically predicted that spontaneous flexoelectric and flexomagnetic effects can emerge in nanoferroic systems.\cite{Eliseev2009} The influence of flexomagnetoelectric interaction on magnons has also been discussed.\cite{Pyatakov2009,Zvezdin2009}
A net magnetization in Mn-based antiperovskite compounds can be induced by flexomagnetic effect.\cite{Lukashev2010}
Spin flexoelectricity, which describes the coupling between bending of magnetization pattern and electric polarization, plays a key role in the emergence of chiral phenomena in magnetic films.\cite{Pyatakov2015}
Rippled GdPtSb membranes exhibit a spontaneous magnetic moment at room temperature driven by flexomagnetism.\cite{Du2021}
Experimental and theoretical studies have reported flexomagnetic effects in Cr$_2$O$_3$ and Sr$_2$IrO$_4$ thin films, as well as in bilayer CrI$_3$.\cite{Makushko2022,Liu2024,Qiao2024}
Moreover, flexomagnetic effects can be significantly enhanced in synthetic antiferromagnets via Dzyaloshinskii-Moriya interaction and the high deformability of skyrmions.\cite{Liu2022} 
A large flexomagnetic response has also been predicted in monolayer CrN, enabled by strain gradient manipulation of the Dzyaloshinskii-Moriya interaction\cite{Gong2025}. 
Strain-gradient-induced skyrmions have been reported in the van der Waals magnet Fe$_3$GaTe$_2$.\cite{Jin2025}
Tuning of magnetocaloric effect in monolayer CrN via flexomagnetism has been also proposed.\cite{Gong2025b}

Since strain gradients are even under time reversal and odd under spatial inversion, the flexomagnetic effect is predominantly indirect. Nevertheless, it has been demonstrated that curvature alone can induce two effective magnetic interactions in curved magnets: an effective magnetic anisotropy and a Dzyaloshinskii-Moriya-like interaction.\cite{Sheka2015,Streubel2016,Edstrom2022,Kuznetsov2025}

In this paper, we discuss the mechanism of flexo Dzyaloshinskii-Moriya interaction. Previous works on this issue mostly rely on phenomenological modelings where contributions from electrons are implicitly or empirically included. The microscopic understanding of this effect is still missing. 
Here, we study interaction between two magnetic impurities on top of curved magnets mediated by itinerant electrons based on a microscopic model treated within perturbation theory.
We show that Dzyaloshinskii-Moriya type interaction can arise from inhomogeneous spin texture by bending. 
Analytical expressions of the Dzyaloshinskii-Moriya type interaction are obtained. We demonstrate this effect in a one-dimensional ring model.
Our mechanism works without any spin-orbit coupling, in contranst to the original Dzyaloshinskii-Moriya interaction\cite{Dzyaloshinsky,Moriya}.

\section{Formulation}

\begin{figure}[htb]
\begin{center}
\includegraphics[clip,width=8.0cm]{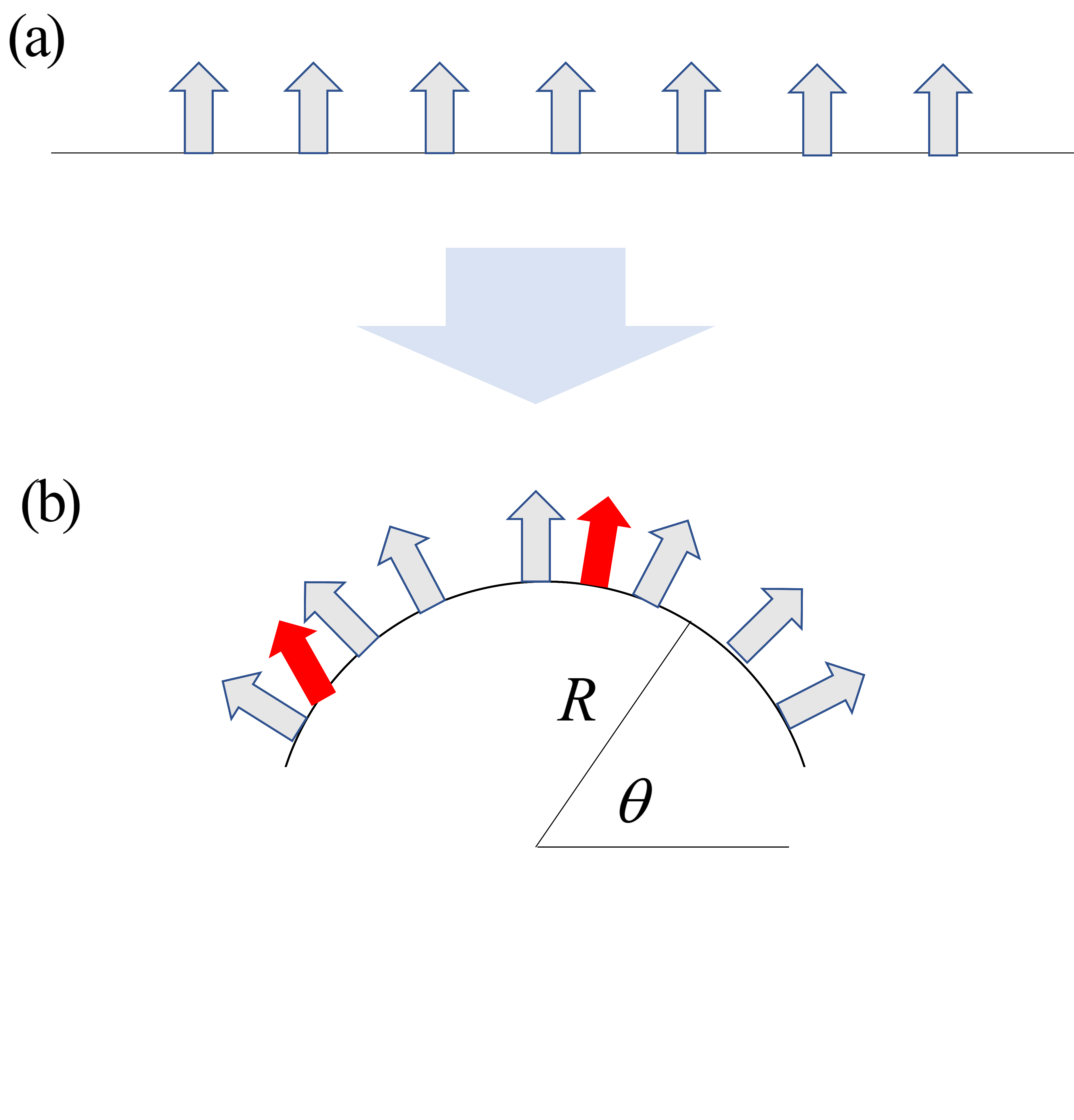}
\end{center}
\caption{
 Uniform magnetization (a) can be made inhomogeneous by bending (b). Two magnetic impurities (red arrows) are placed on the surface of a bent magnet and interact with each other via itinerant electrons.}
\label{f1}
\end{figure}

We consider two localized spins under inhomogeneous background spins which can be realized by bending as shown in Fig. \ref{f1}. The Hamiltonian $H$ is composed of three parts $H_0$, $H_1$ and $H_2$:
\begin{eqnarray}
{H_0} = \frac{{{\hbar ^2}}}{{2m}}{\nabla ^2},\\ {H_1} = \lambda \sum\limits_{i = 1,2} \delta  \left( {{\mathbf{r}} - {{\mathbf{R}}_i}} \right){{\mathbf{S}}_i} \cdot {\bm{\sigma }},\\ {H_2} = J{\mathbf{n}}({\mathbf{r}}) \cdot {\bm{\sigma }}\end{eqnarray}
where $H_0$, $H_1$ and $H_2$ denote the kinetic energy of itinerant electrons, the interaction between the electrons and two local spins, ${{\mathbf{S}}_1}$ and  ${{\mathbf{S}}_2}$, and that between itinerant electrons and background spins described by ${\mathbf{n}}({\mathbf{r}})$. Two spins correnspond to impurity spins. $\bm{\sigma}$ denotes the Pauli matrix in spin space.

The spin-spin interaction between $S_1$ and $S_2$ can be derived by perturbative expansion with respect to $V=H_1+H_2$ to the free energy as follows\cite{RKKY,Kogan,Oriekhov}.
The total free energy is given by \cite{Altland,Nagaosa}
\begin{equation}
F=T \operatorname{Tr} \ln \left[-g^{-1}+V\right]
\end{equation}
which can be expanded with respect to $V$ as
\begin{equation}
F = T \operatorname{Tr} \ln \left[-g^{-1}\right]  +T \operatorname{Tr} \sum_n \frac{1}{n}\left[(gV)^n \right].
\end{equation}
Here, $g$ is the unperturbed Green function associated with $H_0$ and $\operatorname{Tr}$ means the integration or summation over coordinates, Matsubara frequences (or energy) and $\sigma$ space. Note that the localized spins $S_1$ and $S_2$ are treated as external parameters and hence the free energy depends on these spins.
Since we are interested in two spin interaction between $S_1$ and $S_2$, we expand the free energy to the second order in $H_1$.

\subsection{First order in $J$} 

We now investigate the contributions from first order terms in $J$ as shown in Fig. \ref{f2}. By taking trace in $\sigma$ space, they can be calculated as 
\begin{widetext}
\begin{align}
\text{Tr } \mathbf{S}_1 \cdot \boldsymbol{\sigma} \, g(\mathbf{R}_1 - \mathbf{r}) \, \mathbf{n}(\mathbf{r}) \cdot \boldsymbol{\sigma} \, g(\mathbf{r} - \mathbf{R}_2) \, \mathbf{S}_2 \cdot \boldsymbol{\sigma} \, g(\mathbf{R}_2 - \mathbf{R}_1) + \mathbf{S}_1 \cdot \boldsymbol{\sigma} \, g(\mathbf{R}_1 - \mathbf{R}_2) \, \mathbf{S}_2 \cdot \boldsymbol{\sigma} \, g(\mathbf{R}_2 - \mathbf{r}) \, \mathbf{n}(\mathbf{r}) \cdot \boldsymbol{\sigma} \, g(\mathbf{r} - \mathbf{R}_1) \\
 = 2i \left( (\mathbf{S}_1 \times \mathbf{n}(\mathbf{r})) \cdot \mathbf{S}_2 + (\mathbf{S}_1 \times \mathbf{S}_2) \cdot \mathbf{n}(\mathbf{r}) \right) g(\mathbf{R}_1 - \mathbf{r}) g(\mathbf{r} - \mathbf{R}_2) g(\mathbf{R}_2 - \mathbf{R}_1)= 0.
\end{align}
Therefore, we find that no Dzyaloshinskii-Moriya type interaction arises.

\begin{figure}[htb]
\begin{center}
\includegraphics[clip,width=8cm]{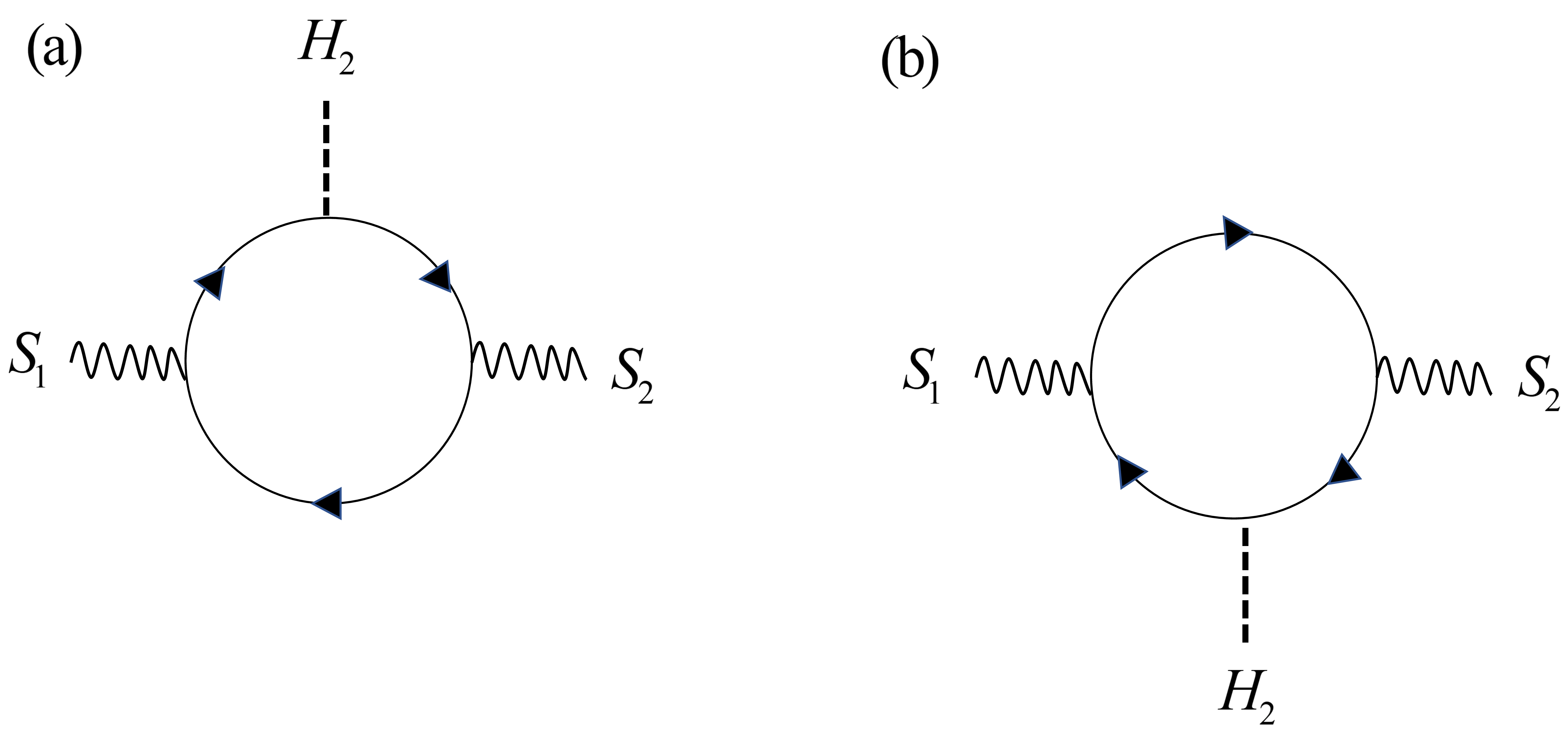}
\end{center}
\caption{
The diagrammatic representations of the free energy to first order in $H_2$.
Solid, wavy and dotted lines represent  electron Green's function, $H_1$ and $H_2$, respectively.}
\label{f2}
\end{figure}

\subsection{Second order in $J$} 

\begin{figure}[htb]
\begin{center}
\includegraphics[clip,width=8cm]{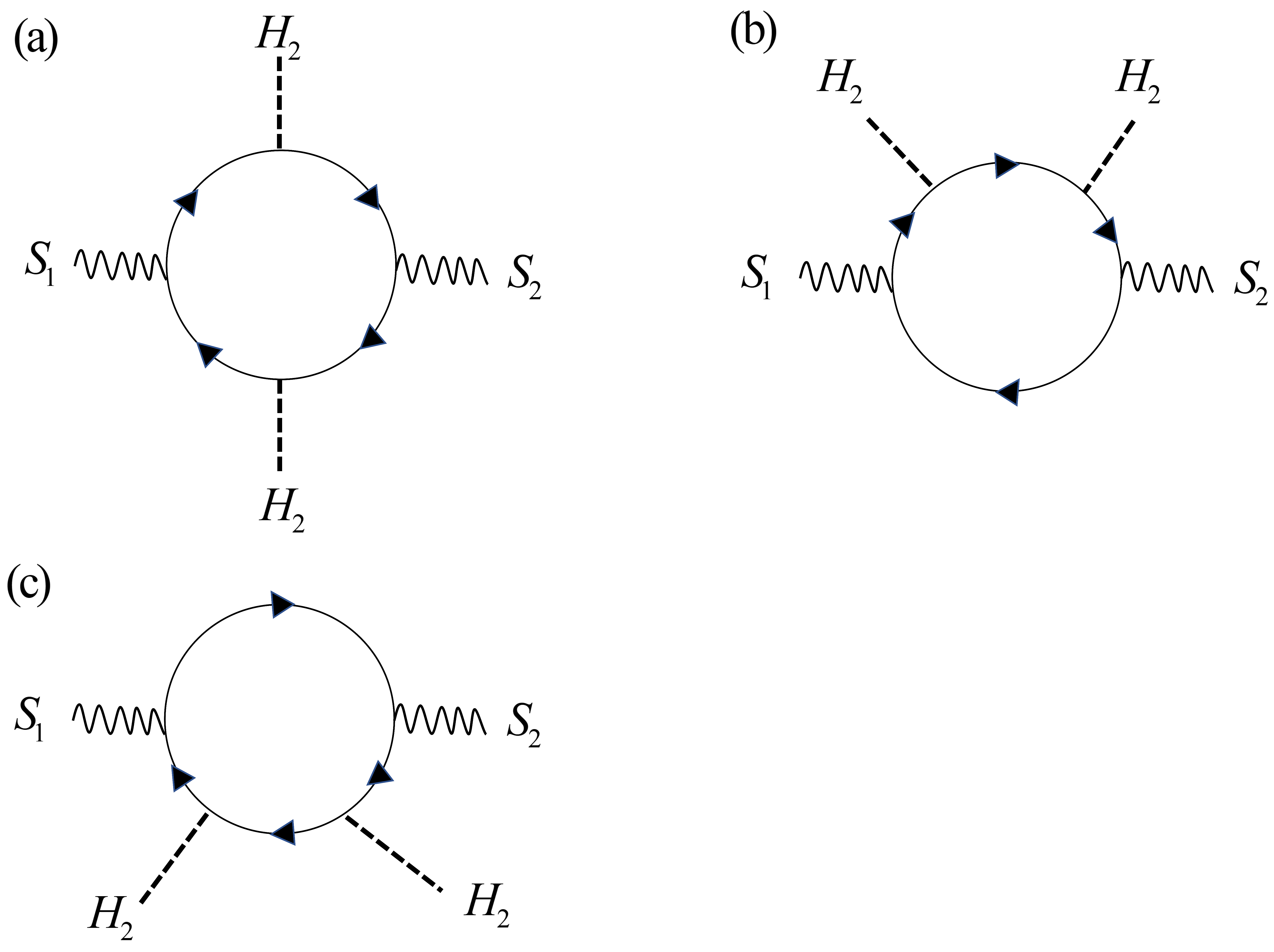}
\end{center}
\caption{
The diagrammatic representations of the free energy to secont order in $H_2$.
Solid, wavy and dotted lines represent  electron Green's function, $H_1$ and $H_2$, respectively.  }
\label{f3}
\end{figure}

We then investigate the contributions from second order terms in $J$ as shown in Fig. \ref{f3}. First, let us consider the contribution from Fig. \ref{f3}(a).  By taking trace in $\sigma$ space, it can be calculated as 
\begin{align}
    &\text{Tr} \left( \mathbf{S}_1 \cdot \boldsymbol{\sigma} \, g(\mathbf{R}_1 - \mathbf{r}) \, \mathbf{n}(\mathbf{r}) \cdot \boldsymbol{\sigma} \, g(\mathbf{r} - \mathbf{R}_2) \, \mathbf{S}_2 \cdot \boldsymbol{\sigma} \, g(\mathbf{R}_2 - \mathbf{r}') \, \mathbf{n}(\mathbf{r}') \cdot \boldsymbol{\sigma} \, g(\mathbf{r}' - \mathbf{R}_1) \right) \\
    &= \text{Tr} \left( \mathbf{S}_1 \cdot \mathbf{n}(\mathbf{r}) + i(\mathbf{S}_1 \times \mathbf{n}(\mathbf{r})) \cdot \boldsymbol{\sigma} \right) \left( \mathbf{S}_2 \cdot \mathbf{n}(\mathbf{r}') + i(\mathbf{S}_2 \times \mathbf{n}(\mathbf{r}')) \cdot \boldsymbol{\sigma} \right) g(\mathbf{R}_1 - \mathbf{r}) g(\mathbf{r} - \mathbf{R}_2) g(\mathbf{R}_2 - \mathbf{r}') g(\mathbf{r}' - \mathbf{R}_1) \\
    &= 2 \left( (\mathbf{S}_1 \cdot \mathbf{n}(\mathbf{r})) (\mathbf{S}_2 \cdot \mathbf{n}(\mathbf{r}')) - (\mathbf{S}_1 \times \mathbf{n}(\mathbf{r})) \cdot (\mathbf{S}_2 \times \mathbf{n}(\mathbf{r}') \right) g(\mathbf{R}_1 - \mathbf{r}) g(\mathbf{r} - \mathbf{R}_2) g(\mathbf{R}_2 - \mathbf{r}') g(\mathbf{r}' - \mathbf{R}_1)
\end{align}
This is symmetric under the exchange of $S_1$ and $S_2$. Thus, there is no Dzyaloshinskii-Moriya type interaction.

Next, let us consider the contributions from Fig. \ref{f3} (b-c). By taking trace in $\sigma$ space, they can be calculated as 
\begin{align}
\text{Tr } \mathbf{S}_1 \cdot \boldsymbol{\sigma} \, g(\mathbf{R}_1 - \mathbf{r}) \, \mathbf{n}(\mathbf{r}) \cdot \boldsymbol{\sigma} \, g(\mathbf{r} - \mathbf{r}') \, \mathbf{n}(\mathbf{r}') \cdot \boldsymbol{\sigma} \, g(\mathbf{r}' - \mathbf{R}_2) \mathbf{S}_2 \cdot \boldsymbol{\sigma} \, g(\mathbf{R}_2 - \mathbf{R}_1) \nonumber \\ + \mathbf{S}_1 \cdot \boldsymbol{\sigma} \, g(\mathbf{R}_1 - \mathbf{R}_2) \, \mathbf{S}_2 \cdot \boldsymbol{\sigma} \, g(\mathbf{R}_2 - \mathbf{r}') \, \mathbf{n}(\mathbf{r}') \cdot \boldsymbol{\sigma} \, g(\mathbf{r}' - \mathbf{r}) \mathbf{n}(\mathbf{r}) \cdot \boldsymbol{\sigma} \, g(\mathbf{r} - \mathbf{R}_1) \\
   = 4 \left( (\mathbf{S}_1 \cdot \mathbf{S}_2) (\mathbf{n}(\mathbf{r}) \cdot \mathbf{n}(\mathbf{r}'))  + (\mathbf{S}_1 \times \mathbf{S}_2) \cdot (\mathbf{n}(\mathbf{r}) \times \mathbf{n}(\mathbf{r}')) \right) g(\mathbf{R}_1 - \mathbf{r}) g(\mathbf{r} - \mathbf{r}') g(\mathbf{r}' - \mathbf{R}_2) g(\mathbf{R}_2 - \mathbf{R}_1)
\end{align}

Thus, the free energy is given by 
\begin{align}
\delta F = \lambda^2 J^2 T \sum \left( \mathbf{S}_1 \times \mathbf{S}_2 \right) \cdot \left( \mathbf{n}(\mathbf{r}) \times \mathbf{n}(\mathbf{r}') \right)  g(\mathbf{R}_1 - \mathbf{r}) g(\mathbf{r} - \mathbf{r}') g(\mathbf{r}' - \mathbf{R}_2) g(\mathbf{R}_2 - \mathbf{R}_1)
\end{align}
where the sum over $\mathbf{r}, \mathbf{r}'$ and the Matsubara frequencies should be taken. This term represents an interaction between two vector spin chiralities.
Hence, we obtain the Dzyaloshinskii-Moriya type interaction of the form
\begin{align}
    \mathbf{D} &= \frac{1}{2} \lambda^2 J^2 T \sum_{} (\mathbf{n}(\mathbf{r}) \times \mathbf{n}(\mathbf{r}')) \{ g(\mathbf{R}_1 - \mathbf{r}) g(\mathbf{r} - \mathbf{r}') g(\mathbf{r}' - \mathbf{R}_2) g(\mathbf{R}_2 - \mathbf{R}_1) - (\mathbf{r} \leftrightarrow \mathbf{r}') \} \\
    &= \frac{\lambda^2 J^2}{2} T \sum_{} (\mathbf{n}(\mathbf{r}) \times \mathbf{n}(\mathbf{r}')) g(\mathbf{R}_2 - \mathbf{R}_1) g(\mathbf{r} - \mathbf{r}')  \{ g(\mathbf{R}_1 - \mathbf{r}) g(\mathbf{r}' - \mathbf{R}_2) - g(\mathbf{R}_1 - \mathbf{r}') g(\mathbf{r} - \mathbf{R}_2) \}.
\end{align}
It has been demonstrated that the Dzyaloshinskii-Moriya interaction stems from  spin-orbit coupling.\cite{Moriya}  Here, we find that the vector spin chirality $\mathbf{n}(\mathbf{r}) \times \mathbf{n}(\mathbf{r}')$ plays a role of spin-orbit coupling.
Eq. (15) also indicates that inhomogeneous background spins in flat magnets induce an effective Dzyaloshinskii-Moriya-type interaction.

\section{1D ring model} 

We now apply the above general results to a one-dimensional (1D) ring model as  shown in Fig. \ref{f1}(b).
We consider a 1D ring of radius $R$ where we use a polar coordinate expression with the angle $\theta$ as illustrated in Fig. \ref{f1}(b), and $S_1$ and $S_2$ are located at $\theta_1$ and $\theta_2$, respectively.

The Green function then reads
\begin{align}
g(l,\theta) = \sum_n \frac{e^{in\theta}}{\varepsilon_l - \frac{\hbar^2 n^2}{2mR^2}} \equiv \sum_n e^{in\theta} A_{l,n}, \; \varepsilon_l=i \omega_l+ E_F
\end{align}
where $\omega_l$ and $E_F$ denote the Matsubara frequency and the Fermi energy, respectively, and $n$ runs over all integers. Hereafter, we set $A_{l,n}=A_{n}$ and omit arguments $l, \theta$ for notational simplicity.
We assume a radial spin texture, i.e., spins perpendicular to the magnet, due to the out-of-plane easy axis in monolayer.\cite{Edstrom2022}
Then, using $\mathbf{n}(\theta) = (\cos\theta, \sin\theta, 0)^T$, the cross product is given by $\mathbf{n}(\theta) \times \mathbf{n}(\theta') = (0, 0, \sin(\theta' - \theta))^T$.
The $z$-component $D_z$ is calculated as:
\begin{align}
    D_z &= \frac{1}{2} \lambda^2 J^2 T \sum \sin(\theta' - \theta) A_{n_1} A_{n_2} A_{n_3} A_{n_4} \cdot e^{in_1(\theta_2 - \theta_1)} e^{in_2(\theta - \theta')}  [ e^{in_3(\theta_1 - \theta)} e^{in_4(\theta' - \theta_2)} - e^{in_3(\theta_1 - \theta')} e^{in_4(\theta - \theta_2)} ]
\end{align}
Integrating over $\theta $ and $\theta' $, we obtain
\begin{align}
D_z = \lambda^2 J^2 T \sum_{l,n_1, n_2} \cos n_1 \theta_{21} \sin(n_2+1)\theta_{21} A_{n_1} A_{n_2} A_{n_2 + 1}^2
\end{align}
with $\theta_{21} = \theta_2 - \theta_1$.
We now define
\begin{align}
D_z = \frac{\lambda^2 J^2} {E_R^3} F(\theta_{21},T,E_F), \; F(\theta_{21}, T,E_F)=T E_R^3 \sum_{l,n_1, n_2} \cos n_1 \theta_{21} \sin(n_2+1)\theta_{21} A_{n_1} A_{n_2} A_{n_2 + 1}^2
\end{align}
with $E_R=\frac{\hbar^2}{2m R^2}$.
Let us estimate this coefficient. For $\lambda \sim J \sim E_R \sim 10$meV, corresponding to $R \sim 1$nm, we obtain $\lambda^2 J^2/E_R^3 \sim$ 10meV.

We show the dependence of $F(\theta_{21}, T, E_F)$ on $\theta_{21}$ for  $E_F/E_R=100$ with $T/E_R=0.25$ and 2.5 in Figs. \ref{f4}(a) and (b), respectively.
It shows an oscillatory behavior as a function of $\theta_{21}$ and increases at low temperature. 

Let us approximate $D_z$. $|A_l|$ takes maxmum at $l=0$ and $n \sim \sqrt{E_F/E_R}=k_F R (\gg 1)$ with $E_F=\hbar^2 k_F^2/(2m)$. Thus, for large $T/E_R$ and $E_F/E_R$, we evaluate the sums in Eq.(18) approximately by this value of $A_l$. Then, we obtain
\begin{align}
D_z \sim \frac{\lambda^2 J^2 }{2\pi^4 T^3} \sin2k_FR \theta_{21}.
\end{align}
We find a RKKY like oscillation.\cite{RKKY} 
Regarding the magnitude of the coefficient, for $\lambda \sim J \sim 10$ meV and $T \sim$ 25 meV, we obtain $\lambda^2 J^2/(2\pi^4 T^3) \sim$ 3 $\mu$eV.

\begin{figure}[htb]
\begin{center}
\includegraphics[clip,width=8cm]{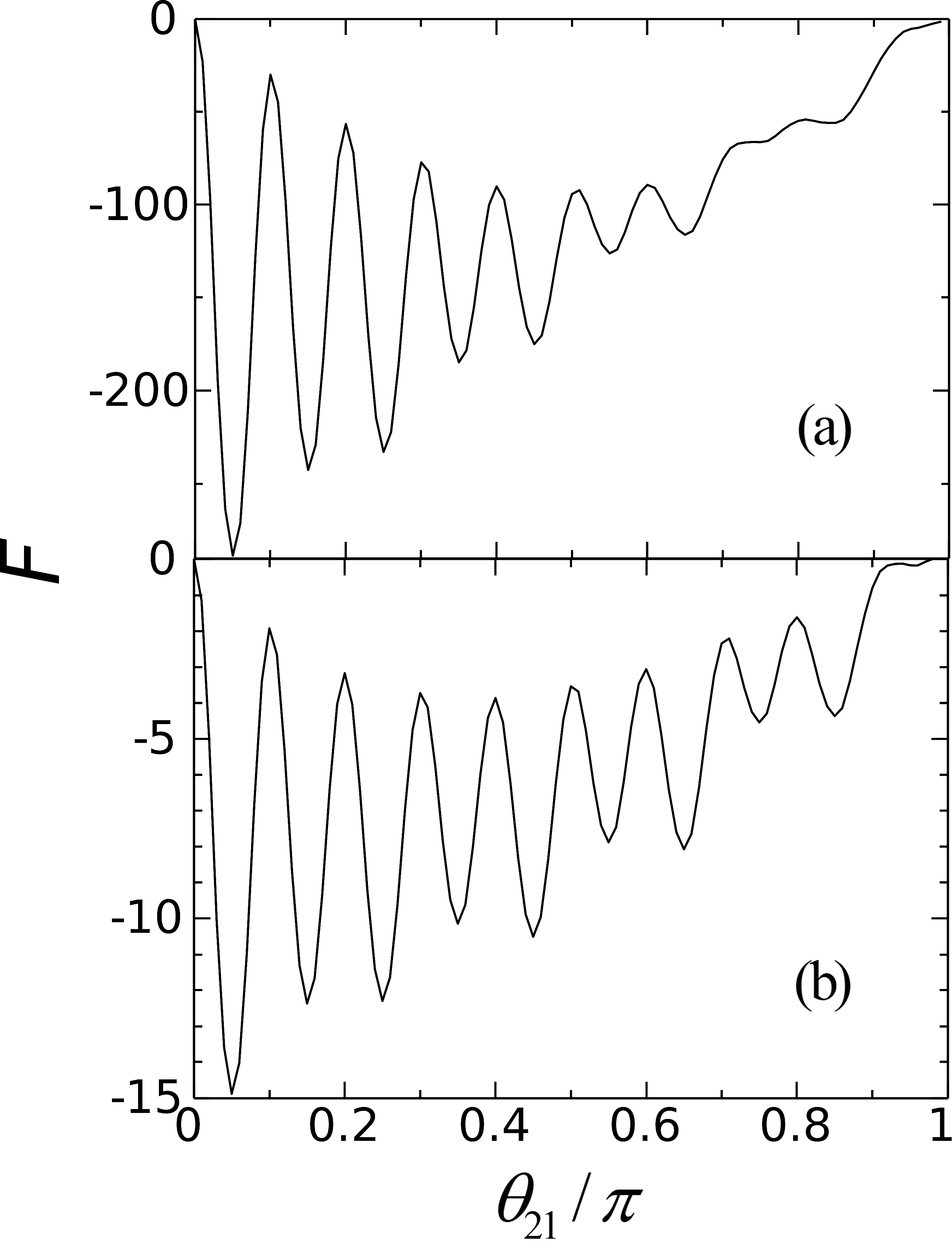}
\end{center}
\caption{
$F(\theta_{21}, T, E_F)$ as a function of $\theta_{21}$ for  $E_F/E_R=100$ with  (a) $T/E_R=0.25$ and (b) 2.5.  }
\label{f4}
\end{figure}

\section{Discussion} 

Our results are also applicable to two-dimensional metallic bent paramagnets with axial phonons. 
Axial (or chiral) phonons are circularly polarized vibrational motions of ions in solids, which can exhibit effective magnetic fields perpendicular to the rotational motion.\cite{ZhangNiu2015,Juraschek2025}
These magnetic fields induce a Zeeman coupling between chiral phonons and spins, which leads to spin-spin interaction via axial phonons. \cite{Yokoyama2024}
Hence, in order to clarify the Dzyaloshinskii-Moriya type interaction mediated by axial phonons, let us replace Eq.(3) by the Zeeman coupling between phonon angular momentum and spin of the form 
\begin{eqnarray}
{H_2} = J'{\mathbf{L}}({\mathbf{r}}) \cdot {\bm{\sigma }}
\end{eqnarray}
where ${\mathbf{L}}({\mathbf{r}})$ denotes the phonon  angular momentum at ${\mathbf{r}}$.
Following the same procedure as in Sec. II, we obtain the following Dzyaloshinskii-Moriya type interaction 
\begin{align}
\mathbf{D} = \frac{\lambda^2 J'^2}{2} T \sum_{} (\mathbf{L}(\mathbf{r}) \times \mathbf{L}(\mathbf{r}')) g(\mathbf{R}_2 - \mathbf{R}_1) g(\mathbf{r} - \mathbf{r}')  \{ g(\mathbf{R}_1 - \mathbf{r}) g(\mathbf{r}' - \mathbf{R}_2) - g(\mathbf{R}_1 - \mathbf{r}') g(\mathbf{r} - \mathbf{R}_2) \}.
\end{align}
This interaction  arises without any magnetism and spin-orbit coupling.
The induced magnetic field can reach 1T \cite{Luo2023} for which we have $J' |\mathbf{L}| \sim 0.1$ meV.

\end{widetext}

\section{Conclusion} 

In conclusion, 
we have developed the microscopic theory of interaction between two magnetic impurities mediated by itinerant electrons on the surface of curved magnets based on perturbation theory. We show that Dzyaloshinskii-Moriya type interaction can arise from inhomogeneous spin texture by bending, without any spin-orbit coupling. 
Analytical expressions of the Dzyaloshinskii-Moriya type interaction are obtained. We have demonstrated this effect in a one-dimensional ring model.

Candidate materials to test our theory are two-dimensional van der Waals ferromagnets such as Cr$_2$Ge$_2$Te$_6$, CrI$_3$, and Fe$_3$GeTe$_2$. \cite{Gong2017,Huang2017,Deng2018}
For example, the exchange interaction in Fe$_3$GeTe$_2$ is on the order of 10 meV, whereas the Dzyaloshinskii-Moriya interaction is about 1 meV.\cite{Macy2021} Thus, the flexo Dzyaloshinskii-Moriya-type interaction predicted in this paper can be dominant in Fe$_3$GeTe$_2$ nanostructures.
The flexo Dzyaloshinskii-Moriya-type interaction can be measured by spin-polarized scanning tunneling microscopy of van der Waals ferromagnetic nanotubes as demonstrated in \cite{Zhou2010}.
One can distinguish the Dzyaloshinskii-Moriya interaction from the flexo Dzyaloshinskii-Moriya-type interaction by considering chirality of crystals. The Dzyaloshinskii-Moriya interaction has opposite signs for opposite crystal chiralities whereas the flexo Dzyaloshinskii-Moriya-type interaction is independent of crystal chirality since it purely stems from spin configurations (irrespective of crystal chirality). Thus, comparing total Dzyaloshinskii-Moriya interactions for left- and right-handed van der Waals ferromagnetic chiral nanotubes, one can extract the flexo Dzyaloshinskii-Moriya-type interaction.

This work was supported by JSPS KAKENHI Grant No.~JP30578216.

\end{document}